# Improvements in Micro-CT Method for Characterizing X-ray Monocapillary Optics

Zhao Wang[a, b], Kai Pan[a, b], Zelin Du[a, b], Shuang Zhang[a, b], Zhiguo Liu[a, b, c, *]

[a]College of Nuclear Science and Technology, Beijing Normal University, Beijing 100875, China

[b]Department of Physics, Applied Optics Beijing Area Major Laboratory, Beijing Normal University, Beijing 100875, China

[c]Beijing Radiation Center, Beijing 100875, China

[*]Corresponding author at: College of Nuclear Science and Technology, Beijing Normal University, Beijing 100875, China.

E-mail address: liuzhiguo@bnu.edu.cn

**Abstract:** Accurate characterization of the inner surface of X-ray monocapillary optics (XMCO) is of great significance in X-ray optics research. Compared with other characterization methods, the micro computed tomography (micro-CT) method has its unique advantages but also has some disadvantages, such as a long scanning time, long image reconstruction time, and inconvenient scanning process. In this paper, sparse sampling was proposed to shorten the scanning time, GPU acceleration technology was used to improve the speed of image reconstruction, and a simple geometric calibration algorithm was proposed to avoid the calibration phantom and simplify the scanning process. These methodologies will popularize the use of the micro-CT method in XMCO characterization.

## 1 Introduction

X-ray monocapillary optics (XMCO) based on the principle of total external reflection show excellent performance, and are a hot spot in X-ray optics research [1-2]. To satisfy the modulation requirements of the X-ray propagation path in different applications, XMCO can be designed into different shapes such as tapered [3], ellipsoid [4], and parabolic [5]. At present, XMCO has been widely used in various X-ray optical systems such as microscale X-ray fluorescence [6], diffraction [7], imaging [8], and small-angle scattering [9].

According to the principle of total external reflection, the shape design of XMCO is the key, and the slope error will make the inner surface of XMCO deviate from the expectation, thus affecting the effect of X-ray modulation [3-5]. Accurate characterization of the inner surface of XMCO can not only evaluate its performance but also improve the processing technology as feedback, which is of great significance in X-ray optics research [10]. Currently, there are two methods to measure the inner diameter of XMCO i.e. indirect and direct. The indirect method measures the outer diameter of XMCO and estimates the inner diameter on the assumption that the ratio of the inner diameter to outer diameter is constant [10-11]. Due to the flow and cooling of the molten glass in the glass tube drawing process, the internal and external diameter ratio of XMCO is not fixed, and the accuracy of the indirect method is limited [5, 10]. The direct method uses advanced measuring methods such as optical microscopy [12], scanning electron microscopy [13], three-dimensional confocal microscopy [14], X-ray imaging [15], and micro-CT [5] to directly measure the inner diameter of XMCO. Among these instruments, micro-CT has unique advantages in that it can nondestructively provide a 3D structure of XMCO and has the potential to measure multiple XMCOs simultaneously. Micro-CT is a cone beam CT using micro focus X-ray source and high-resolution detector. By collecting X-ray attenuation information from multiple views and using appropriate image reconstruction algorithm, the internal structure of the sample can be reconstructed. According to the spatial resolution of the system, micro-CT can be further divided into two categories: conventional micro-CT (resolution: 50 ~ 100 microns) and high-resolution micro-CT (resolution: 1 ~ 20 microns). At present, micro-CT has been widely used in the field of life science and material science, due to its high resolution [16]. However, micro-CT measurement also has some shortcomings. For example, scanning and 3D tomographic image reconstruction both consume a significant amount of time, and the commonly used geometric calibration method based on calibration phantoms needs to scan phantoms before sampling, which is inconvenient.

To better apply micro-CT in the characterization of XMCO, this paper will shorten the time consumption of scanning and image reconstruction and simplify the scanning process. The main innovations described in this paper include the following: 1) sparse sampling is used to shorten the scanning time, 2) GPU acceleration technology is used to shorten the tomographic image reconstruction time, and 3) a projection image self-correcting (PISC) algorithm is proposed to avoid the use of calibration phantoms and simplify the scanning process.

## 2 Method
### 2.1 Algorithms and Implement
#### 2.1.1 Sparse sampling and incomplete data reconstruction algorithm

Sparse sampling begins with medical CT because sparse sampling can reduce the radiation dose to the human body. Although micro-CT has no strict limit on radiation dose, sparse sampling is also very attractive because micro-CT usually needs to sample hundreds or thousands of projections to ensure high resolution, and a full scan takes a long time. Sparse sampling can not only shorten the scanning time but also reduce the measurement error caused by power fluctuation and focus shift of the X-ray tube [17]. In the case of incomplete projection data caused by sparse sampling, traditional image reconstruction algorithms such as FDK and SART have poor image quality and cannot meet actual needs, and it is necessary to choose an appropriate incomplete data reconstruction algorithm for image reconstruction. As a classic type of incomplete data reconstruction algorithm, the TV minimization algorithm based on compressed sensing is widely used. The TV minimization algorithm assumes that the pixel values of a tomographic image are piecewise constants, and an abrupt change in pixel values usually occurs only on the boundary of the internal structure. XMCO has the characteristics of piecewise constants, so this paper will use the TV minimization algorithm to reconstruct the tomographic image of XMCO in the case of undersampling.

The POCS-TV algorithm proposed by Sidky et al. [18] is representative of the TV minimization algorithm. The mathematical expression of POCS-TV is as follows:

$$\min_{\vec{f}} \|\vec{f}\|_{TV} \text{ s.t } \mathbf{A}\vec{f} = \vec{p}, f_j \geq 0$$
$$\|\vec{f}\|_{TV} = \sum_{i,j} \sqrt{\left(f_{i+1,j} - f_{i,j}\right)^2 + \left(f_{i,j+1} - f_{i,j}\right)^2} \quad (1)$$

In formula (1), $\vec{f}$ is the discrete image vector, $\|\vec{f}\|_{TV}$ is the total variation of $\vec{f}$, $\mathbf{A}$ is the system matrix that models the CT scanning process as a linear system, and $\vec{p}$ is the measured projection data. The POCS-TV algorithm uses a hybrid strategy to solve the constrained minimization problem. That is, the gradient descent method is used to minimize the total variation, the POCS method is used to impose data consistency and positive constraints, and the iteration is repeated until the algorithm converges. In Section 3.3, we will use the POCS-TV algorithm to reconstruct the tomographic image of XMCO in the undersampling case.

**2.1.2 GPU accelerated projection/backprojection algorithm**

Compared with the conventional CT system, the resolution of micro-CT is higher, which means that the computational burden of micro-CT image reconstruction is heavier and the computing time is longer. Reducing the time consumption of the image reconstruction process can improve the efficiency of characterizing XMCO. Therefore, it is of great significance to improve the reconstruction speed of micro-CT. Many studies have shown that GPU acceleration technology is an effective method for meeting this requirement [19-20].

The time-consuming projection/backprojection is a bottleneck in the process of image reconstruction. Fortunately, projection/backprojection also has a high degree of parallelism. In voxel-driven [21] or ray-driven [22-23] methods, the operation for each voxel or ray is the same and independent. The GPU used to accelerate the projection/backprojection greatly improves the speed of image reconstruction. Moreover, reasonable application of hardware-accelerated texture memory can further improve the efficiency for a large amount of interpolation in projection/backprojection.

In this paper, we implement some common algorithms such as FDK, SART, and POCS-TV. Their GPU-projection/backprojection kernels are programmed based on NVIDIA's Compute Unified Device Architecture (CUDA), which is one of the most popular GPU parallel computing platforms.

**2.1.3 PISC algorithm**

In this paper, we propose a projection image self-correcting (PISC) algorithm based on the Radon transform. PISC algorithm can directly solves the key geometric errors from the projection image of XMCO.

Geometric calibration is the key to ensuring the measurement accuracy of micro-CT. Generally, the geometric errors of micro-CT system can be attributed to the relative displacement and rotation of the detector, and can be divided into four categories, as shown in Fig. 1: tilt, slant, skew, and offset [24]. It should be noted that for the sake of the drawing effect, Fig. 1 magnifies these errors, the rotation error is only approximately 1 degree, and the offset error is only a few pixels. The most common geometric calibration method of the micro-CT system is the offline method, which calculates the geometric errors according to the relationship between a calibration phantom and its projection. The offline calibration method needs to scan the phantom before scanning the sample, which brings inconvenience to the scanning process of micro-CT. In addition, because the field of view (FOV) of the micro-CT system is usually very small, the machining accuracy of the phantom must be very high, which also makes the phantom expensive. Our PISC algorithm avoids the use of phantoms, simplifies the scanning process of micro-CT, and reduces the cost.

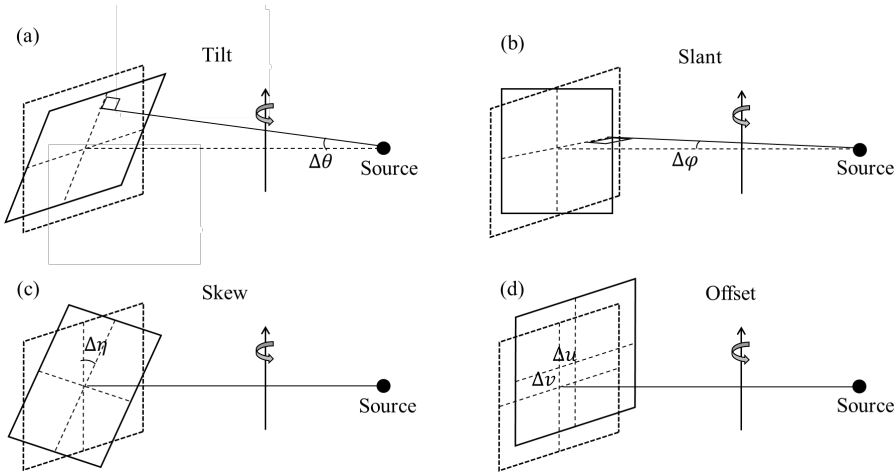

**Fig. 1** Schematic diagram of geometric error of micro-CT system

(Dotted box indicates ideal detector position; solid box indicates detector position with error)

After considering the influence of each geometric error on XMCO characterization, the PISC algorithm only calibrates the key errors $\Delta\eta$ and $\Delta u$ and ignores the minor errors $\Delta\theta$, $\Delta\varphi$, and $\Delta v$. It is popular to simplify the calibration process by reasonably ignoring some geometric errors [24-27]. We ignore $\Delta\theta$ and $\Delta\varphi$ because a large number of studies have shown that $\Delta\theta$ and $\Delta\varphi$ have little impact on the quality of image reconstruction and can be safely ignored [24, 27]. Moreover, due to the small FOV of micro-CT, multisegment projection is often merged, which makes up for the impact of $\Delta v$. It is worth noting that although there are also errors in the measurement of the distance between the X-ray source and detector (SDD) and the distance between the X-ray source and rotation axis (SOD), $\Delta SDD$ and $\Delta SOD$ can be safely ignored for practical considerations. The dimensionless quantities of interest in XMCO characterization, such as the slope and the ratio of the inner diameter to outer diameter, are not affected by $\Delta SDD$ and $\Delta SOD$.

PISC algorithm calibrates $\Delta\eta$ and then $\Delta u$. The principle of $\Delta\eta$ calibration is as follows: if the Radon transform is applied to binary XMCO projection images from different rotation angles, then only when the rotation angle is $\Delta\eta$ does the number of zero elements in the corresponding transformation result reach the maximum. The skew can be corrected by rotating the XMCO projection image by $-\Delta\eta$. After correcting the skew, the principle of $\Delta u$ calibration is also very simple: according to the symmetry of XMCO, if the center column index of nonzero elements is calculated row by row for a binary XMCO projection image, then the number of pixels from the average of the index to the center column of the detector is $\Delta u$. The lateral offset can be corrected by translating the XMCO projection image by $-\Delta u$ pixels.

## 2.2 Experiment

Fig. 2 shows the experimental platform and ellipsoidal XMCO sample used in this paper. The microfocus X-ray source uses a copper target. The focus size is 8 μm when the power is 4 W. The model of the rotary stage is Suruga Seiki KPW06360, and the step angle accuracy can reach 0.004 degree. The model of the CCD flat panel detector is Hamamatsu C11440-22CU, and the effective detection area is 13.3×13.3 mm$^2$, including 2048×2048 pixels. The scanning parameter configuration of the micro-CT system is shown in Table 1.

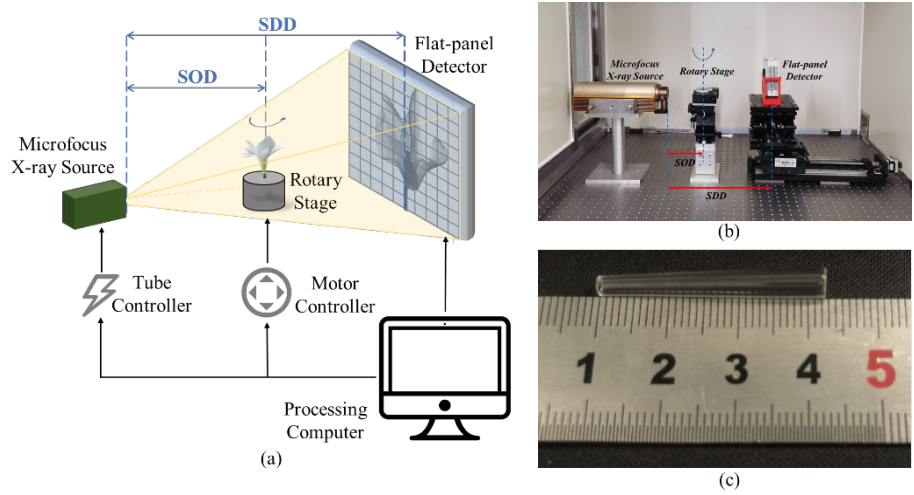

**Fig. 2**  (a) Micro-CT scanning schematic diagram; (b) desktop micro-CT system; (c) ellipsoidal XMCO sample

The information obtained by the flat-panel detector is the number of X-ray photons arriving at each pixel during the exposure time. By simply transforming the Lambert-Beer law, the X-ray intensity can be converted into the line integral of the attenuation coefficient as follows:

$$p = \int_0^L \mu(x)dx = \log \frac{I_0 - I_d}{I - I_d} \qquad (2)$$

In formula (2), $p$ is the line integral of the X-ray attenuation coefficient (called the projection), $I$ is the X-ray intensity after sample attenuation, $I_0$ is the bright background, and $I_d$ is the dark background. The number of X-ray photons recorded by the detector follows a Poisson distribution. In this paper, the average of three measurements for $I$ and five measurements for $I_0$ and $I_d$ are taken to reduce the noise level. In addition, a median filter with a window width of 7×7 is used to convolute the projection image to further smooth the noise. Compared with other low-pass filters (such as the mean filter and Gaussian filter), the median filter has the advantage of maintaining a sharp edge. Fig. 3 shows the detector output signal, bright background, dark background, and projection image calculated according to formula (2).

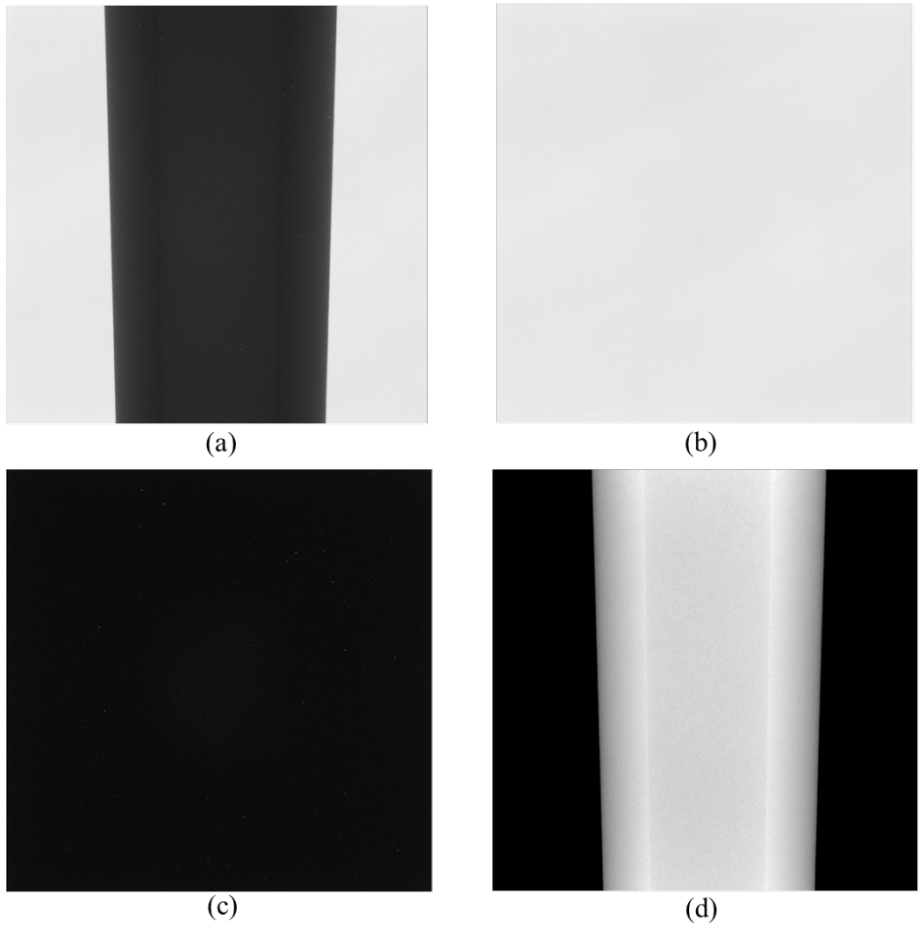

**Fig. 3** (a) Detector output signal at 1 degree; (b) bright background; (c) dark background; (d) projection image at 0 degrees

Table 1　Scanning parameter configuration of micro-CT system in this paper

| Scanning Parameter Configuration | |
|---|---|
| Voltage (kV) | 40 |
| Current (mA) | 0.1 |
| Exposure Time (s) | 20 |
| SDD (mm) | 343 |
| SOD (mm) | 251 |
| FOV (mm) | 6×6×6 |
| Detector Pixels | 2048×2048 |
| Detector Size (mm) | 13.3×13.3 |
| Sampling Range (degrees) | 0 ~ 180 |
| Sampling Interval (degrees) | 1 |

## 3 Results
### 3.1 Speedup of GPU acceleration

The most direct way to evaluate the effect of GPU acceleration is to compare the execution time of single GPU-projection/backprojection and CPU-projection/backprojection. The speedup can be defined as follows:

$$\text{speedup} = \frac{\text{Time to perform single CPU-projection/backprojection}}{\text{Time to perform single GPU-projection/backprojection}} \quad (3)$$

To evaluate the effect of GPU acceleration under different computational loads, the time consumption of single GPU-projection/backprojection and single CPU-projection/backprojection are measured under three typical parameter settings, as shown in Fig. 4. To reduce the measurement error, each group of data given in Fig. 4 is the average value calculated after one warm-up and five repeated measurements. It is necessary to explain that this study is carried out on a laptop with a 64-bit Windows 10 system. The CPU of the laptop is a 1.60 GHz i5-8265U, and the RAM is 16 GB. The GPU of the laptop is a GeForce MX250 (Pascal architecture) with a maximum clock frequency of 1.58 GHz, three SMs (384 CUDA cores in total), and 2 GB of memory.

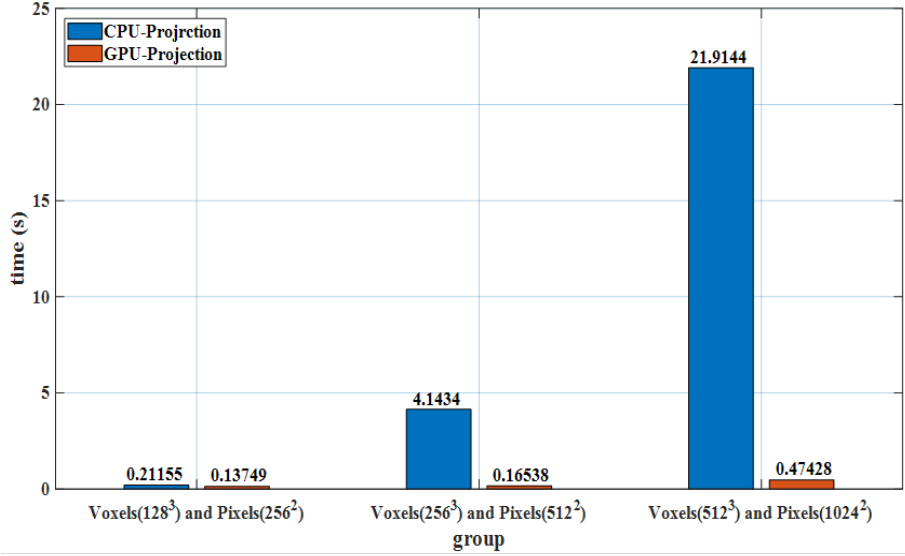

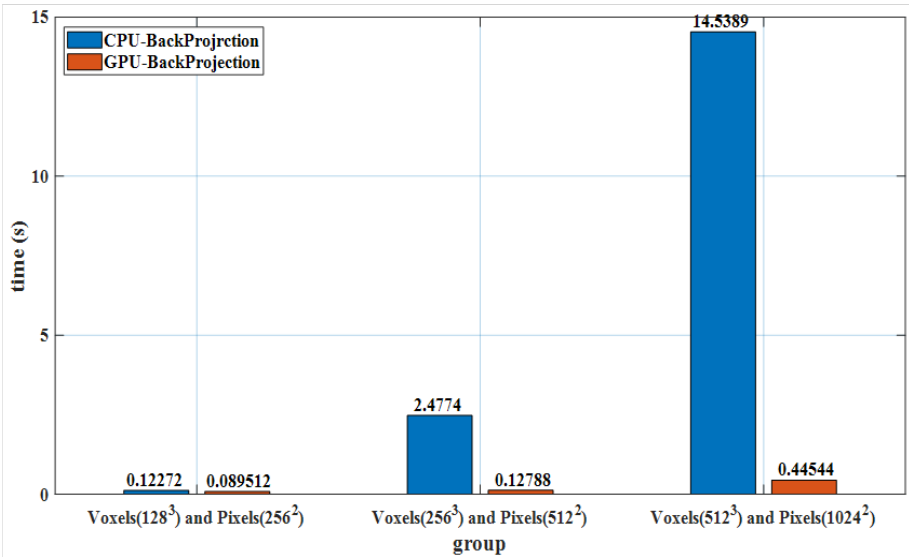

**Fig. 4** (a) Comparison of time consumption of projection; (b) comparison of time consumption of backprojection

With an increase in the computational load, the GPU acceleration effect becomes significant: in the case of 512×512×512 voxels and 1024×1024 pixels, the speedup of GPU-projection/backprojection can reach 46.2 and 32.6, respectively. A larger computational load means that the time-consuming data transmission between the CPU and GPU can be better hidden by the execution time of the kernel, and the advantages of GPU acceleration can be better reflected. Generally, the acceleration effect of a GPU is related to its hardware architecture and the number of CUDA cores. In this paper, in the case of 512×512×512 voxels, 1024×1024 pixels, and 360 sampling angles, the reconstruction time of the analytical reconstruction algorithm is less than 1 minute, and that of the iterative reconstruction algorithm is usually dozens of minutes.

## 3.2 Result of PISC algorithm

The performance of the PISC algorithm mainly depends on the sensitivity of skew calibration. This paper will test the sensitivity of skew calibration at small $\Delta\eta$ by numerical simulation. According to the shape of the XMCO projection image, we designed the digital phantom shown in Fig. 5(a) as the ground truth. By rotating Fig. 5(a) to a given $\Delta\eta$ and adding 1% Gaussian white noise, the test image shown in Fig. 5(b) can be obtained. This will be used as the input of the PISC algorithm. In the numerical simulation, a total of 21 groups of $\Delta\eta$ were tested from -5 to 5 degrees at an interval of 0.5 degree. The angle search step of the Radon transform was set to 0.01 degree, and the simulation was repeated 10 times. Fig. 5(c) shows one of the outputs of the PISC algorithm, and Fig. 5(d) shows the relationship between the given and calibrated $\Delta\eta$. It is obvious that the skew calibration is very accurate even for a small $\Delta\eta$ of approximately 1 degree.

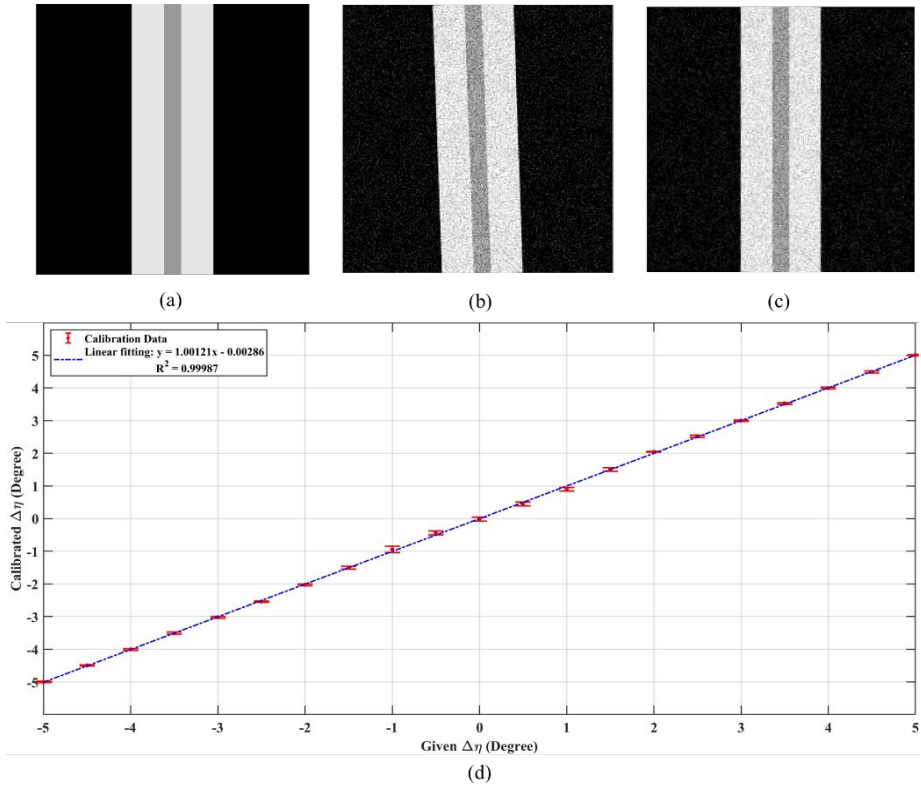

**Fig. 5** (a) Ground truth; (b) test image; (c) output of PISC algorithm;

(d) Relationship between given and calibrated $\Delta\eta$

To illustrate the effectiveness of the PISC algorithm, tomographic images reconstructed from raw projections and PISC-corrected projections are compared, as shown in Fig. 6. Here, the image reconstruction algorithm we used is SART (10 iterations). Fig. 6(a) shows a "false contour", which is caused by skew and lateral offset, while Fig. 6(b) shows a closed contour and clear edge. The PISC algorithm can effectively eliminate the geometric artifacts caused by skew and lateral offset.

In this paper, the maximum values of $\Delta\eta$ and $\Delta u$ calibrated by the PISC algorithm are 0.84 degrees and 28 pixels, respectively.

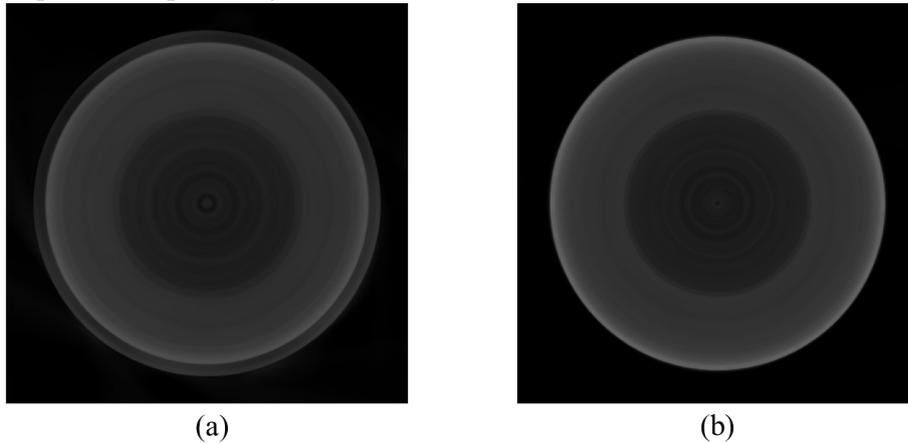

(a)　　　　　　　　　　　　　　　(b)

**Fig. 6**　Transverse section of XMCO reconstructed from (a) raw projections and (b) PISC-corrected projections (window level is 0.5, window width is 1)

## 3.3 Result of sparse sampling and POCS-TV algorithm

Sparse sampling can not only shorten the scanning time but also reduces the measurement error caused by the power fluctuation and focus shift of the X-ray tube. In this section, the POCS-TV algorithm is used to reconstruct the tomographic image of XMCO under different numbers of sampling views, and the relationship between the quality of the reconstructed image and the number of sampling views is studied to reveal the feasibility of sparse sampling applied to XMCO. The reconstructed image of the SART algorithm in the case of full-sampling will be used as the benchmark to evaluate the reconstructed image of the POCS-TV algorithm in the case of undersampling, and the quantitative metric we used is the RMSE, which is defined as follows:

$$\text{RMSE} = \frac{\left\|\vec{f} - \vec{f}^*\right\|_2}{N} \qquad (4)$$

In formula (4), $\vec{f}$ and $\vec{f}^*$ represent the reconstructed images of POCS-TV and SART respectively, and $N$ is the number of voxels of the 3D image (in this paper, $N = 512^3$). The smaller the RMSE, the closer $\vec{f}$ and $\vec{f}^*$ are.

Table 2 and Table 3 show the parameters used for SART and POCS-TV respectively, and the introduction of these parameters can be found in [18]. Here, we give the best parameters selected by repeated tests, which can give the best visual results. Fig. 7 shows the trend of RMSE with the number of sampling views, and the iterations of both SART and POCS-TV are 50. The RMSE tends to be stable when the number of sampling views exceeds 50. Fig. 8 compares the reconstructed images of SART (full-sampling, 180 views) and POCS-TV (undersampling, 50 views). The images are visually very similar, meaning that high-quality XMCO tomographic images can be reconstructed with only 50 sampling views. Fig. 9 is a three-dimensional visualization results of the XMCO image reconstructed by POCS-TV, which is obtained by Avizo software.

Table 2    Parameters used for SART algorithm

| Parameters Used for SART Algorithm | |
|---|---|
| relaxation parameter | 1.2 |

Table 3    Parameters used for POCS-TV algorithm

| Parameters Used for POCS-TV Algorithm | |
|---|---|
| relaxation parameter | 1.2 |
| balance parameter | 0.2 |
| number of gradient descent steps | 20 |

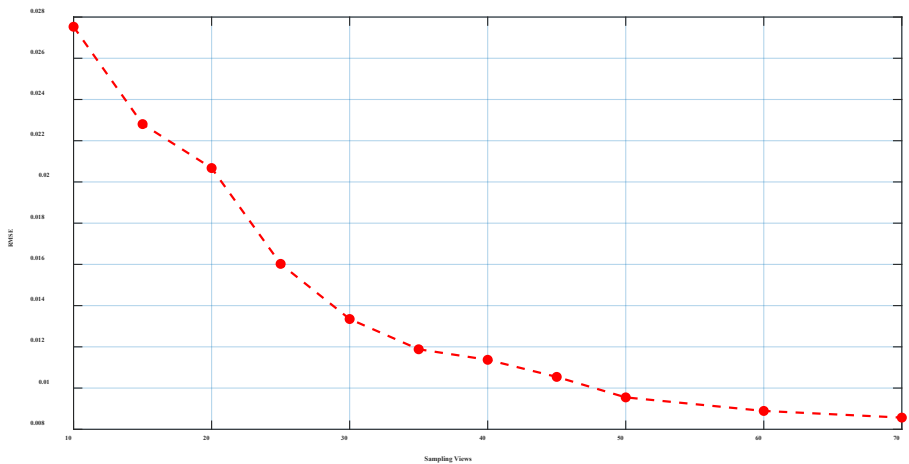

**Fig. 7**   Trend of RMSE with number of sampling views

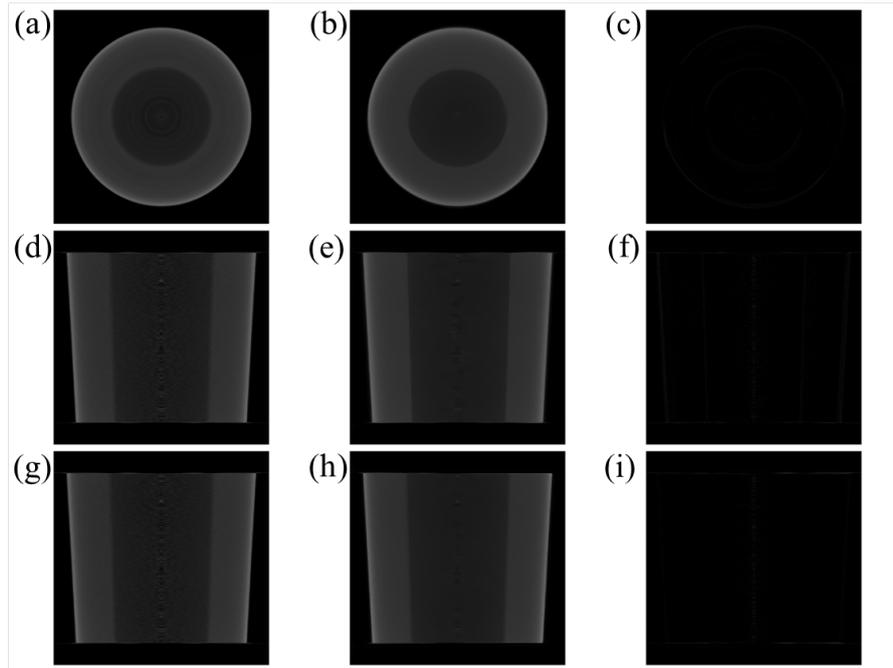

**Fig. 8** (a), (d), and (g) Transverse, sagittal, and coronal sections of the reconstructed image of SART; ~~(d)~~(b), (e), and (h) Transverse, sagittal, and coronal sections of the reconstructed image of POCS-TV; (c), (f), and (i) Transverse, sagittal, and coronal section of residual image

(window level is 0.5, window width is 1)

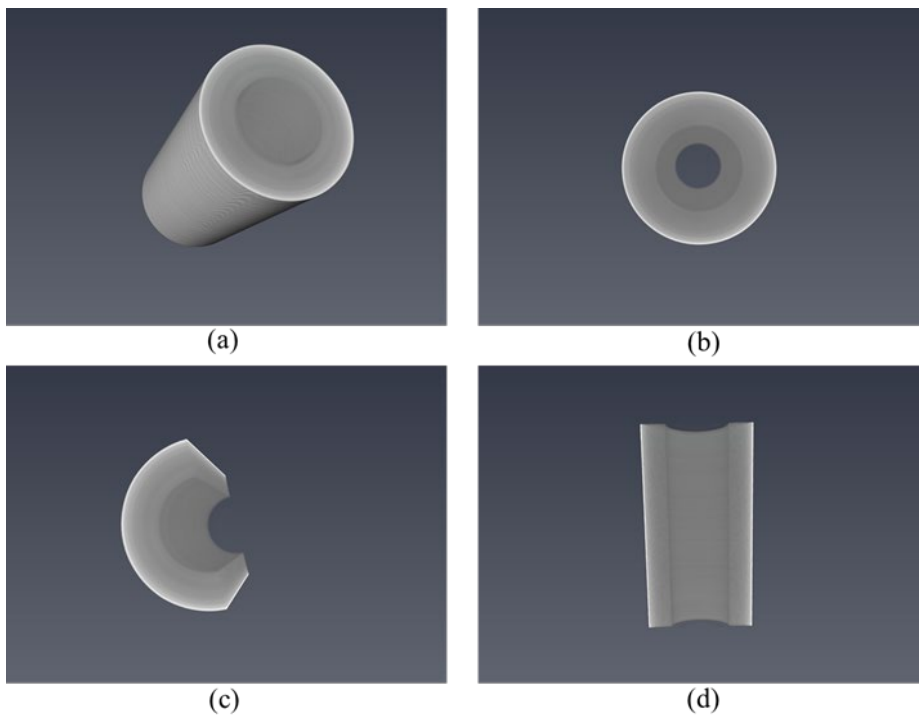

**Fig. 9** Three-dimensional visualization of the XMCO image reconstructed by POCS-TV: (a) surface morphology; (b) transverse section; (c) sagittal section; (d) coronal section

Furthermore, according to the reconstructed image of POCS-TV (undersampling, 50 views), some slope data of XMCO can be calculated in increments of $\Delta = 0.2\,\text{mm}$. As a comparison, the outer diameter of the same XMCO is measured by a laser rangefinder (LRF) with $\Delta = 0.2\,\text{mm}$ as the moving step, and the slope curve of XMCO is obtained by ellipse fitting. Here, we use equation $ax^2 + bxy + cy^2 + dx + ey + f = 0$ to fit the ellipse, and the fitting technique can refer to [28]. The fitting parameters we calculated are shown in Table 3, and the goodness of fit is $R^2 = 0.9873$. As shown in Fig. 10, the slope data calculated by the micro-CT method are in good agreement with the slope curve fitted by the LRF method, indicating that sparse sampling is feasible.

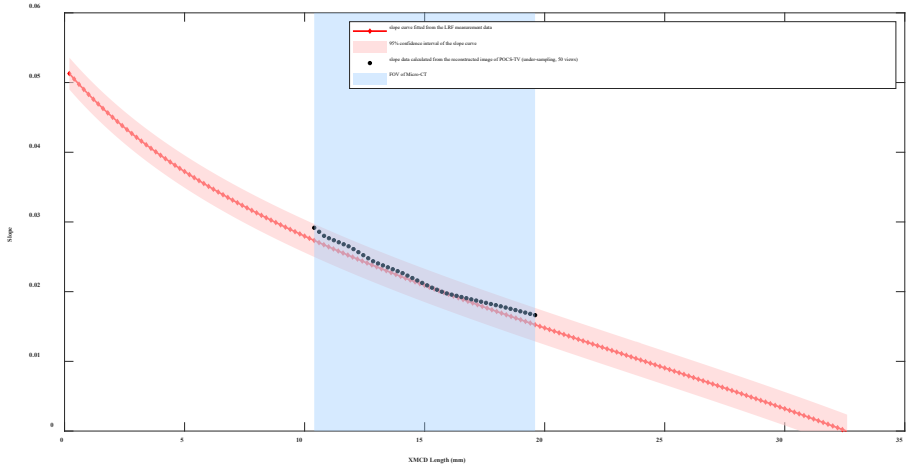

**Fig. 10** Comparison between slope data and slope curve

**Table 4** Fitting parameters of elliptic equation

| Equation: $ax^2 + bxy + cy^2 + dx + ey + f = 0$ | |
|---|---|
| a | 1.1982e-04 |
| b | 0.0018 |
| c | 0.0858 |
| d | -0.0213 |
| e | -0.3362 |
| f | 0.9376 |

# 4 Discussion and Conclusion

The inner surface characteristics of XMCO directly affect its performance in modulating X-rays, so it is important to characterize the inner surface accurately. Compared with other characterization methods, the micro-CT method makes it possible to evaluate the 3D structure of XMCO on the micron scale nondestructively and has the potential to measure multiple XMCOs simultaneously. At present, micro-CT can achieve spatial resolution at the micron level, which is very attractive for the characterization of XMCO. However, higher resolution often means longer scanning time and image reconstruction time.

In this paper, XMCO was sparsely sampled, and then the POCS-TV algorithm was used to reconstruct tomographic images from incomplete projection data to reduce the micro-CT scanning time. We found that high-quality XMCO tomographic images can be reconstructed using only 50 sampling views in one-third of the sampling time compared to full sampling using 180 sampling views. To further evaluate the feasibility of using sparse sampling in the micro-CT method for XMCO characterization, slope characterization was taken as an example. By comparing the slope data calculated by the reconstructed image of POCS-TV and the slope curve fitted by the LRF method, we found that they are in good agreement. It is worth noting that the minimum sampling views required for high-quality image reconstruction are related to the incomplete data reconstruction algorithm used. If more advanced algorithms, such as dictionary learning [28] and low-dimensional manifold models [30] are used, then the scanning time can be further shortened because these algorithms apply more advanced mathematical models to describe the sparse property of XMCO.

Although sparse sampling can shorten the scanning time and reduce the measurement errors related to scanning time such as power fluctuation and focus shift of the X-ray tube, the iterative nature of the incomplete data reconstruction algorithm also makes the image reconstruction process very time-consuming. To improve the speed of image reconstruction, this paper used GPU acceleration technology to accelerate the projection/backprojection operation involved in the image reconstruction algorithm and successfully control the time consumption of the POCS-TV algorithm within tens of minutes. Both the GPU hardware and the acceleration strategy we used are not optimal, and the discussion of state-of-the-art GPU acceleration research is beyond the scope of this paper. What we have done is only a reference, readers can use more advanced hardware and acceleration strategies to achieve a shorter time consumption of image reconstruction.

In this paper, the PISC algorithm was proposed to further popularize the use of the micro-CT method. The PISC algorithm is an online geometric correction method that automatically corrects key geometric errors and carries out corresponding geometric corrections according to the projection of XMCO. Because the PISC algorithm does not depend on a calibration phantom, it simplifies the scanning process and lowers the purchase cost of the phantom. In addition, the PISC algorithm is an analytical algorithm in nature, so it requires very little computation compared with other online calibration methods based on objective function optimization. It should be noted that the PISC algorithm is not a universal geometric calibration method and is only applicable to the case where the scanned object can be approximated to a simple convex shape.

In conclusion, sparse sampling can greatly reduce the time consumption of micro-CT scanning and effectively suppress the time-related errors caused by the power fluctuation and focus shift of the X-ray tube. GPU acceleration can drastically improve the speed of image reconstruction. The PISC algorithm can accurately calibrate key geometric errors without the calibration phantom, which effectively simplifies the scanning process. These methodologies will popularize the use of the micro-CT method in XMCO characterization.